\documentclass[aps,prl,twocolumn,showpacs,10pt,floatfix]{revtex4-1}

\usepackage{amsmath}
\usepackage{graphics}
\usepackage{dcolumn}

\input{MyMc}
\newcommand{\ExH}[2]{#2}

\newboolean{ShowSec}%without backslash
\setboolean{ShowSec}{false}%<-- change here
\newcommand{\Sec}[1]{\ifthenelse{\boolean{ShowSec}}{\section{#1}}{}}
\newcommand{\SubSec}[1]{\ifthenelse{\boolean{ShowSec}}{\subsection{#1}}{}}
\newif\ifPRL
\PRLtrue

\setboolean{ShowCorrections}{false}

\begin{document}

%\title{The butterfly effect in a point-like phase battery}
%\title{The butterfly effect in a point-like $\varphi$ Josephson junction}
\title{Phase retrapping in a point-like $\varphi$ Josephson junction: the butterfly effect}

\author{E. Goldobin}
\author{R. Kleiner}
\author{D. Koelle}
\affiliation{%
  Physikalisches Institut and Center for Collective Quantum Phenomena in LISA$^+$,
  Universit\"at T\"ubingen, Auf der Morgenstelle 14, D-72076 T\"ubingen, Germany
}

\author{R.G. Mints}
\affiliation{%
  The Raymond and Beverly Sackler School of Physics and Astronomy,
  Tel Aviv University, Tel Aviv 69978, Israel
}

\date{%
  \today\ File: \textbf{\jobname.\TeX}
}

\begin{abstract}
  We consider a $\varphi$ Josephson junction, which has a bistable zero-voltage state with the stationary phases $\psi=\pm\varphi$. In the non-zero voltage state the phase ``moves'' viscously along a tilted periodic double-well potential. When the tilting is reduced quasistatically, the phase is retrapped in one of the potential wells. We study the viscous phase dynamics to determine in which well ($-\varphi$ or $+\varphi$) the phase is retrapped for a given damping, when the junction returns from the finite-voltage state back to zero-voltage state. In the limit of low damping the $\varphi$ Josephson junction exhibits a butterfly effect --- extreme  sensitivity of the destination well on damping. This leads to an impossibility to predict the destination well.
\end{abstract}

\maketitle

\Sec{Introduction}

%\footnote{The name originates from the title of the famous talk of Edward N. Lorenz, the discoverer of the effect, ``Predictability: does the flap of the butterfly's wings in Brazil set off a tornado in Texas?''}

The term \emph{butterfly effect} is widely used to denote the extreme sensitivity of complex dynamical systems to initial conditions\cite{Scott:2005:ENLS,Sprott:2003:Chaos}. The effect puts a clear distinction between determinism and predictability. For example, due to the butterfly effect it is not possible to predict the weather reliably for more than 3--5 days in advance. Although the original work\cite{Lorenz:1963:DetNonperFlow} was related to simulation of atmospheric processes, it was discovered later on that in quite a few problems of nonlinear physics a tiny perturbations of initial conditions \ifPRL\else or system parameters\fi might lead to completely different final states. In particular, the butterfly effect was observed in \ifPRL\else numerical\fi simulations of long Josephson junction subjected to an oscillating magnetic field\cite{Yugay:2006:LJJ:Butterfly}.

%Social systems also exhibit the butterfly effect. In the presidential elections in the Palm Beach country, Florida, USA in the year 2000 few questionable ballots caused the chaos and finally made George W. Bush rather than Al Gore the 43rd president of the USA\footnote{From the original publication\cite{Sinclair:2000:Elections} it becomes obvious that it was a double-butterfly effect --- the so-called ``butterfly ballots'' caused the butterfly effect in a sense described here}. The butterfly effect is an necessary condition for chaos in deterministic systems.

\begin{figure}[!tb]
  \includegraphics{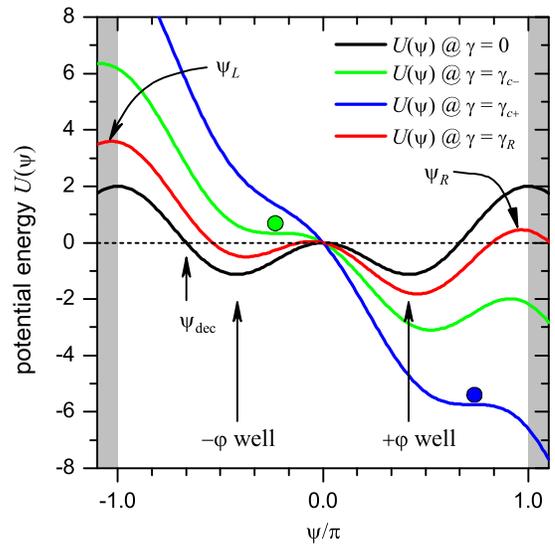}
  \caption{%
    Tilted periodic double well (Josephson) potential  $U(\psi)$ for $\Gamma_0=-4$ and different values of bias current (tilt) $\gamma$.
  }
  \label{Fig:Pot}
\end{figure}

Consider now a point-like $\varphi$ Josephson junction ($\varphi$-JJ) proposed theoretically\cite{Mints:1998:SelfGenFlux@AltJc,Buzdin:2003:phi-LJJ,Goldobin:2011:0-pi:H-tunable-CPR} and recently demonstrated experimentally\cite{Sickinger:2012:varphiExp}. This $\varphi$ JJ has a doubly degenerate ground state phase $\psi=\pm\varphi$, which is a result of the unusual
\iffalse
effective current phase relation (CPR). For short junction \ExH{and small applied magnetic field $h$} the CPR can be represented as
%
\begin{equation}
  I_s(\psi) = \sin(\psi)+\frac{\Gamma_0}{2}\sin(2\psi)
  \ExH{+ \Gamma_h h \cos(\psi)}
  , \label{Eq:CPR}
\end{equation}
%
where parameter $\Gamma_0<0$ \ExH{and $\Gamma_h$} is defined by asymmetry between 0 and $\pi$ parts\cite{Goldobin:2011:0-pi:H-tunable-CPR,Lipman:varphiEx}. Since $I_s(\psi)=U_J'(\psi)$, this corresponds to the
\fi
Josephson energy profile
\begin{equation}
  U_J(\psi) = 1-\cos(\psi) + \frac{\Gamma_0}{4}[1-\cos(2\psi)] \ExH{+ \Gamma_h h \sin(\psi)}{}
  . \label{Eq:U_J(psi)}
\end{equation}
The energy $U_J(\psi)$ has a form of a $2\pi$-periodic double-well potential with wells at $\psi=\pm\varphi$ \ExH{at $h=0$}{}, see Fig.~\ref{Fig:Pot}. The ground state phase $\varphi=\arccos(-1/\Gamma_0)$ \ExH{at $h=0$}{}. The parameter\ExH{s}{}\ $\Gamma_0<0$ \ExH{and $\Gamma_h$}{} define\ExH{}{s}\ the depth of the wells \ExH{and asymmetry between them}{}\cite{Goldobin:2011:0-pi:H-tunable-CPR,Lipman:varphiEx}. The potential has two wells per period for $\Gamma_0<-1$ \ExH{and not very large $|h|$}{}. \ExH{The normalized magnetic field $h$ changes the asymmetry between the two wells and can make the double well potential single welled for large $|h|$.}{} Application of a bias current $\gamma$ tilts the potential as shown in Fig.~\ref{Fig:Pot}.

Since in the zero-voltage state the $\varphi$-JJ is bistable, it is interesting to understand, in which of these two states the phase is retrapped when the $\varphi$ JJ returns from the finite-voltage state to the zero-voltage state upon quasistatic decrease of the tilt (bias current density) $\gamma$. Note, that for conventional 0 or for $\pi$ JJs with only a single energy minimum per period of Josephson energy, such a question does not arise. Earlier\cite{Goldobin:CPR:2ndHarm} we have na\"ively suggested that upon returning from the positive-voltage state to the zero-voltage state the phase is retrapped in the $+\varphi$ state. This is indeed true for rather large damping\cite{Sickinger:2012:varphiExp}. However, at lower damping the behavior is non-trivial and often experimentally seems to be non-deterministic\cite{Sickinger:2012:varphiExp}.

Here we study the retrapping process of the Josephson phase in a point-like $\varphi$ JJ and demonstrate that at low damping the system exhibits the butterfly effect.

\Sec{Model}

The dynamics of a $\varphi$ JJ can be described by the equation of motion for the phase $\psi(t)$ (see the supplement)
\begin{equation}
  \ddot{\psi} + \fracp{U_J}{\psi} = \gamma - \alpha\dot{\psi}
  , \label{Eq:EffEq}
\end{equation}
where $\Gamma_0<-1$ \ExH{and $\Gamma_h$ are}{is a} parameter\ExH{s}{} of the potential defining its depth \ExH{and asymmetry, accordingly, $h$ is the normalized magnetic field}{}, $\gamma$ is the bias current density normalized to the average critical current density $\av{j_c(x)}$, $\alpha$ is the dimensionless damping coefficient (also normalized using $\av{j_c(x)}$). This model describes well a $\varphi$ JJ made out of a 0-$\pi$ JJ with facet lengths $L_0\lesssim\lambda_{J,0}$ and $L_\pi\lesssim\lambda_{J,\pi}$ \cite{Mints:1998:SelfGenFlux@AltJc,Buzdin:2003:phi-LJJ,Goldobin:2011:0-pi:H-tunable-CPR,Lipman:varphiEx}, where $\lambda_{J,0}$ and $\lambda_{J,\pi}$ are the Josephson lengths in the 0 and $\pi$ parts, accordingly. For 0-$\pi$ JJs with somewhat longer facets, like in experiment\cite{Sickinger:2012:varphiExp}, it holds qualitatively. For experimental parameters the double well potential $U(\psi)$ calculated numerically is not so deep as in the model \eqref{Eq:U_J(psi)} as can be seen in Fig.~1 of Ref.~\onlinecite{Sickinger:2012:varphiExp}. This leads only to quantitative rather than qualitative difference in the results obtained.

Equation~\eqref{Eq:EffEq} describes a phase (point-like particle of a unit mass with the coordinate $\psi$) moving viscously (term $-\alpha\dot{\psi}$) in a tilted $2\pi$-periodic double-well potential
\begin{equation}
  U(\psi) = U_J(\psi) - \gamma\psi
  , \label{Eq:U(psi)}
\end{equation}
see Fig.~\ref{Fig:Pot}. The bias current density $\gamma$ sets the tilt.

\ifPRL\else
Equation~\eqref{Eq:EffEq} also describes a complex pendulum with a deviation angle $\psi$ having two stable positions $\psi=\pm\varphi$.
\footnote{The system can be constructed as two coupled pendula. The first pendulum hangs on the axis 1 with a deviation angle denoted as $\psi$. The second pendulum hangs on axis 2 connected with axis 1 by means of two gears with ratio 2:1 (bigger gear at axis 1, smaller gear at axis 2) so that when axis 1 makes half-turn, the axis 2 makes full-turn. The gears are connected so that when pendulum 1 hangs down, pendulum 2 sticks up. The parameter $|\Gamma_0|$ sets the ratio of the moment of inertia of pendulum 2 relative to pendulum 1. \ExH{To model the effect of the $\Gamma_h$ term one needs a third pendulum connected to the first one with gear ratio 1:1 and shifted by a $\pi/2$ angle.}{}}
\fi

\Sec{Results}

The main process that we are interested in here is the dynamics of switching from the finite-voltage state to the zero-voltage state. At \ExH{$h=0$ and}{} $\gamma=0$ the phase is trapped in one of the wells of the potential $U(\psi)$, \ie, at $\psi=-\varphi$ or at $\psi=+\varphi$. Upon increase of the bias current $\gamma$, the potential $U(\psi)$ tilts and, at some value of the bias current $\gamma$, the phase escapes because the corresponding well disappears. For $\gamma>0$ the phase escapes from the $-\varphi$ well at $\gamma=\gamma_{c-}$ and from the $+\varphi$ well at $\gamma=\gamma_{c+}>\gamma_{c-}$ as found earlier\cite{Goldobin:2011:0-pi:H-tunable-CPR,Sickinger:2012:varphiExp}, see Fig.~\ref{Fig:Pot}. After escape, the phase slides viscously along the periodic potential. The voltage across the junction is proportional to the velocity of the phase motion $\dot{\psi}(t)$. Further, we start decreasing the bias current density (tilt) $\gamma$ \emph{quasistatically}. At some $\gamma=\gamma_R$, which depends of the damping $\alpha$, the phase is retrapped in one of the wells. It is this retrapping process, which is the main subject of this study.

Note that in general the damping $\alpha$ is a function of temperature $T$.
%For tunnel junction $\alpha\propto\exp(-\Delta/k_BT)$, where $\Delta$ is the energy gap.
However, the temperature is also responsible for thermal fluctuations that can be added as an additional stochastic current to the \rhs of Eq.~\eqref{Eq:EffEq}. In the following we assume that such fluctuations are negligible (zero) and the only effect of temperature is the change in $\alpha$. At the end we discuss shortly the effect of these fluctuations on our results.

\SubSec{Retrapping current}

To analyze the retrapping process, first, we search the value of the tilt $\gamma_R$ (retrapping current) at a given damping $\alpha$. The retrapping situation corresponds to the trajectory, on which the phase starts with zero velocity at the main maximum of the potential $U(\psi)$ situated at $\psi=\psi_L$, see Fig.~\ref{Fig:Pot}, slides down viscously, passes two minima and one maximum and arrives to $\psi_R=\psi_L+2\pi$ with zero velocity. The value of $\psi_L$ is one of the roots of the equation $\partial U/\partial\psi=0$, \ie, from Eq.~\eqref{Eq:U(psi)},
\begin{equation}
  \left. \fracp{U_J}{\psi} \right|_{\psi=\psi_L} = \gamma
  , \label{Eq:psi_L}
\end{equation}
corresponding to the maximum of $U(\psi)$.

\begin{figure}[!htb]
  \includegraphics{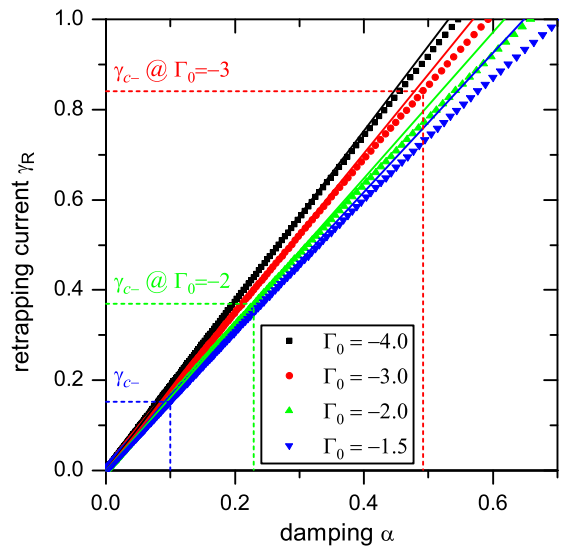}
  \caption{%
    The dependence $\gamma_R(\alpha)$ for different values of $\Gamma_0$. Symbols represent the results of direct numerical simulation. Lines show PT results, given by Eq.~\eqref{Eq:gamma_R(alpha)} for $\alpha\to0$. The horizontal dashed lines show the values of the depinning current $\gamma_{c-}$ for given $\Gamma_0$, \ie, the current, at which the $-\varphi$ well disappears. For $\gamma>\gamma_{c-}$ the potential has only one $+\varphi$ well and the phase is retrapped there. The vertical dashed line shows the corresponding value of $\alpha_R(\gamma_{c-})$. For JJ with $\alpha>\alpha_R(\gamma_{c-})$ the retrapping current $\gamma_R(\alpha)>\gamma_{c-}$ and potential has only one $+\varphi$ well where the phase is retrapped.
  }
  \label{Fig:gamma_c(alpha)}
\end{figure}

Since $\psi_L$ depends on $\gamma$, it is more convenient to fix the tilt $\gamma$ and look for the critical/retrapping value of $\alpha_R(\gamma)$, rather than looking for $\gamma_R(\alpha)$. To find $\psi(t)$ the Eq.~\eqref{Eq:EffEq} was solved for fixed $\gamma$ and $\alpha$ with initial conditions $\psi(0)=\psi_L+\epsilon$ and $\dot{\psi}(0)=0$ (typically we use $\epsilon\sim10^{-6}$). The solution was calculated up to the point where either $\psi(t)>\psi_L+2\pi$ or where $\dot{\psi}<0$. In the first case the particle is \emph{not} trapped for given $\gamma$ and $\alpha$ and moves to the next period of the potential. In the second case the particle is trapped. By varying $\alpha$ we repeat the simulation to find the boundary values $\alpha_R(\gamma)$ between the above two cases with a given accuracy of $10^{-6}$. The resulting plots of already inverted $\gamma_R(\alpha)$ dependences for different values of $\Gamma_0$ are shown in Fig.~\ref{Fig:gamma_c(alpha)}. One can see that the dependences $\gamma_R(\alpha)$ are almost linear.

In the limit of $\alpha\to0$ and $\gamma\to0$ one can use a simple perturbation theory (PT) to obtain the slope of this linear dependence. We assume that $\alpha$ and $\gamma$ are perturbations. Without perturbations ($\alpha=\gamma=0$) \ExH{an at $h=0$}{} the phase dynamics is governed by the equation
\begin{equation}
  \ddot{\psi}+\left[ \sin(\psi) + \frac{\Gamma_0}{2}\sin(2\psi)\right] = 0
  , \label{Eq:Pendulum.NotPert}
\end{equation}
which has the first integral
\begin{equation}
  \dot{\psi}^2 = C + \left[ 2\cos(\psi) + \frac{\Gamma_0}{2}\cos(2\psi)\right]
  , \label{Eq:psi-dot}
\end{equation}
where $C=2-\frac{\Gamma_0}{2}$ is determined from the initial condition for the retrapping trajectory: $\dot{\psi}(-\infty)=0$, $\psi(-\infty)=\psi_L$. Now, if we turn on the small damping $\alpha$ and the bias $\gamma$, they will lead to dissipation and driving, correspondingly. The dissipated energy $Q$ along the ``critical'' path from $\psi_L=-\pi$ to $\psi_L+2\pi=+\pi$ (for $\gamma=0$) is
\begin{equation}
  Q = \alpha \int_{-\pi}^{+\pi} \dot{\psi}\,d\psi = \alpha I(\Gamma_0)
  , \label{Eq:Q}
\end{equation}
where, using Eq.~\eqref{Eq:psi-dot}, we define
\begin{equation}
  I(\Gamma_0) = \int_{-\pi}^{+\pi} \sqrt{2[1+\cos(\psi)] - \Gamma_0\sin^2(2\psi)}\,d\psi
  , \label{Eq:I}
\end{equation}
which can be calculated numerically for any $\Gamma_0$.

The energy input due to the tilt $\gamma$ is
\begin{equation}
  E_\gamma = \gamma \left[\psi_L+2\pi - \psi_L \right] = 2\pi\gamma
  . \label{Eq:E_gamma}
\end{equation}
In the case of retrapping trajectory, $E_\gamma$ compensates $Q$ and brings the particle exactly to the position $\psi_L+2\pi$ with zero velocity. Thus, from $E_\gamma=Q$ we get
\begin{equation}
  \gamma_R(\alpha) = \frac{I(\Gamma_0)}{2\pi}\alpha
  . \label{Eq:gamma_R(alpha)}
\end{equation}
We note that in the limit $\Gamma_0\to0$ we have $I(\Gamma_0)\to8$ and obtain a well known result\cite{Stewart68} valid for conventional JJ with sinusoidal CPR, namely $\gamma_R=\frac4\pi \alpha$. The lines corresponding to $\gamma_R(\alpha)$ dependences \eqref{Eq:gamma_R(alpha)} are shown in Fig.~\ref{Fig:gamma_c(alpha)} and agree well with numerical data for $\alpha\to0$.

\SubSec{Which well? (Destination well)}

\begin{figure}[!htb]
  \includegraphics{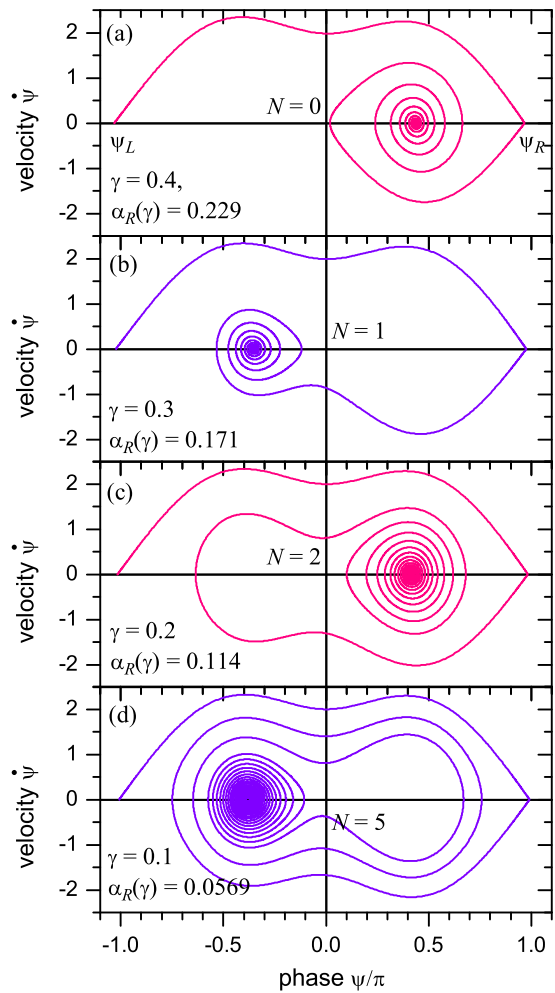}
  \caption{%
    The retrapping trajectories in the phase plane $(\dot{\psi},\psi)$ for $\Gamma_0=-3$ and different tilt $\gamma$. The phase starts at $\psi_L$, where $U(\psi)$ has a maximum, with $\dot{\psi}=0$ and arrives to $\psi_R=\psi_L+2\pi$ (next $U(\psi)$ maximum) with $\dot{\psi}=0$. Then the phase falls back and is trapped in one of the minima of the $U(\psi)$.
  }
  \label{Fig:Traj}
\end{figure}
\begin{figure}[!htb]
  \includegraphics{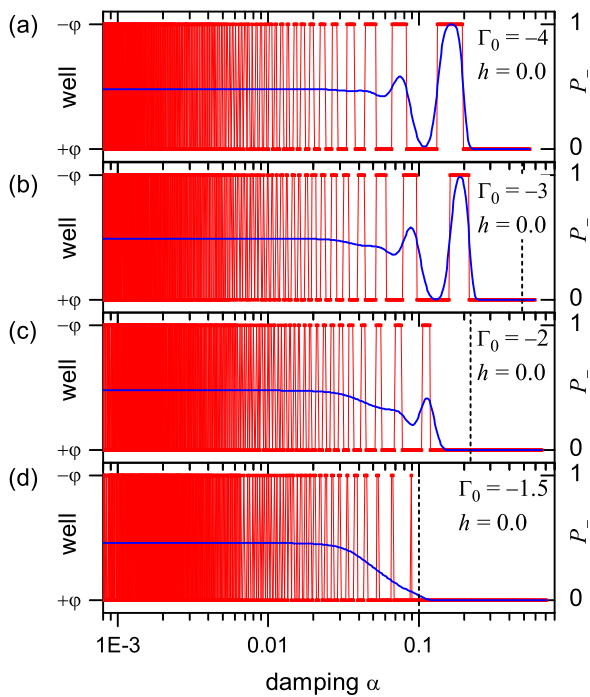}
  \caption{(Color online)
    The well in which the particle is trapped at $\gamma_R(\alpha)$ as a function of $\alpha$ (symbols). Vertical dashed line show the value $\alpha_R(\gamma_{c-})$. For $\alpha>\alpha_R(\gamma_{c-})$ the potential has only a single $+\varphi$ well. In (a) $\alpha_R(\gamma_{c-})\approx1.327$ and is not visible. The lines show the effect of the low frequency Gaussian noise with $\sigma_\gamma=0.02$ in the bias current circuitry. In this case the right vertical axis represents the probability $P_-$ to find the system in the $-\varphi$ state.
  }
  \label{Fig:well-alpha}
\end{figure}

Knowing the dependence $\alpha_R(\gamma)$ we now take various values of $\gamma$, take the corresponding $\alpha_R(\gamma)$, put the phase at the turning point $\psi=\psi_R(\gamma)-\epsilon$, see Fig.~\ref{Fig:Pot}, and follow its time-evolution. The ultimate goal is to see in which well ($-\varphi$ or $+\varphi$) the phase is trapped. The decision about trapping is taken when the velocity $\dot{\psi}$ changes the sign two times in a row on the \emph{same} side relative to the energy barrier separating the two wells. Examples of the trajectories on the phase plane $(\dot{\psi},\psi)$ are shown in Fig.~\ref{Fig:Traj}. Thus we get destination well \vs $\alpha_R(\gamma)$ dependence.

Figure~\ref{Fig:well-alpha} shows the destination well ($-\varphi$ or $+\varphi$) as a function of $\alpha$. One can see that, indeed, for large $\alpha$ the phase is trapped in the $+\varphi$ well, as predicted\cite{Goldobin:2011:0-pi:H-tunable-CPR} and demonstrated experimentally\cite{Sickinger:2012:varphiExp}. However, as $\alpha$ decreases, the destination well changes from $+\varphi$ to $-\varphi$ then back to $+\varphi$ and so on. The intervals of $\alpha$, corresponding to the retrapping in a particular well, become smaller and small even on a logarithmic scale, see Fig.~\ref{Fig:well-alpha}. Thus, in the limit of small $\alpha$ the destination well is \emph{extremely sensitive to the initial conditions} --- tiny variation (or fluctuation) of $\alpha$ or $\gamma$ (thermal or electronic noise) results in a global effect --- retrapping in a different well. Thus our $\varphi$ JJ exhibits the \emph{butterfly effect}.

In our case, the butterfly effect prevents one to forecast, in which well the phase will be retrapped in experiment in the limit of small $\alpha$. In fact, already in the first experimental work\cite{Sickinger:2012:varphiExp} on $\varphi$ JJs it was seen that the retrapping is not deterministic at low damping (temperatures $\sim 300\units{mK}$). In experiment, due to inevitable presence of noise, the destination well \vs $\alpha$ curve will be smeared. If we assume a low frequency Gaussian electronic noise of the amplitude $\sigma_\gamma$ in the bias circuitry, one can calculate the probability $P_-$ to find the system in the $-\varphi$ state by making a convolution of $\pm\varphi(\alpha)$ curve with the Gaussian distribution of width $\sigma_\gamma$. The resulting $P_-(\alpha)$ is also shown in Fig.~\ref{Fig:well-alpha}. One can see that the noise smears the fast switchings and $P_-\to1/2$ at $\alpha\to0$.

However the presented model is oversimplified. If one includes a stochastic (instrumental noise or thermal fluctuations) current term in the \rhs of Eq.~\eqref{Eq:EffEq}, it will also affect the value of the retrapping current $\gamma_R(\alpha)$ making it not well defined (smeared) with the  ensemble average $\av{\gamma_R(\alpha)}$ larger than $\gamma_R(\alpha)$ calculated above\cite{Ben-Jacob:1982:LifeTime,Fenton:2008:JJ:Esc+Retrap}. A rigorous tretment of noise will be presented elsewere.
%\EG{May be he wants us to delete convoluted curves completely from Fig.~\ref{Fig:well-alpha}?}

\iffalse
\old{If the noise is dominated by thermal fluctuations, $\sigma(\alpha)$ decreases with $\alpha$ (temperature $T$) exactly in the same way as the size of the regions (see below) in Fig.~\ref{Fig:well-alpha}. However, at some $\alpha<\alpha^*$ ($T<T^*$) the quantum fluctuations should dominate the dynamics and the averaging similar to the one shown in Fig.~\ref{Fig:well-alpha} with $P_-\to0.5$ at $\alpha\to0$ should take place again.}
\fi

\iffalse
If the noise is dominated by thermal fluctuations due to finite temperature, we have two simultaneous effects when $T\to0$: (i) $\sigma$ of the noise current decreases $\propto T$, and (ii) $alpha$ decreases $\propto \exp(-\Delta/k_BT)$, while the size of the regions (see below) in Fig.~\ref{Fig:well-alpha} decreases as $\Delta\gamma\propto\Delta\alpha\propto\alpha^2$. That is smearing due to fluctuations decreases slower than the region size in Fig.~\ref{Fig:well-alpha} and the oscillations will average out at $T\to0$. Moreover,} at some $\alpha<\alpha^*$ ($T<T^*$) the quantum fluctuations should dominate the dynamics and the (stronger) averaging similar to the one shown in Fig.~\ref{Fig:well-alpha} with $P_-\to0.5$ at $\alpha\to0$ should take place.
\fi

%
\begin{figure}[!t]
  \includegraphics{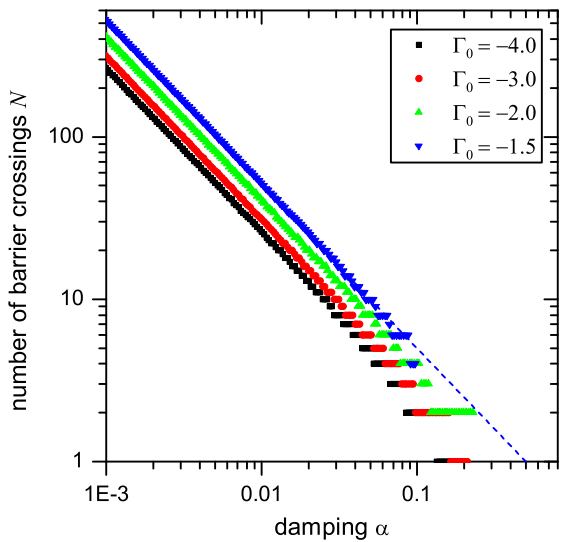}
  \caption{%
    The dependence $N(\alpha)$ for different values of $\Gamma_0$. At low $N$ one sees that $N$ is integer. At high $N$ it is a straight line on this double log-scale, \ie, corresponds to $N\propto 1/\alpha$.
  }
  \label{Fig:N(alpha)}
\end{figure}

At the end we would like to mention an interesting detail. When simulating a retrapping dynamics, we also have counted how many times $N$ the phase crossed the barrier separating the wells before being trapped in one of the wells. In fact the well ($+\varphi$ or $-\varphi$) plotted in Fig.~\ref{Fig:well-alpha} is just $\varphi [1-2(N \bmod 2)]$. Fig.~\ref{Fig:N(alpha)} shows $N(\alpha)$ plots for different values of $\Gamma_0$. Note that $N$ is an integer so it changes step-wise, as it is well visible for large $\alpha$. For small $\alpha$ (large $N$) the dependence looks almost continuous and can be well approximated by $N\approx C_\alpha/\alpha$. The coefficient $C_\alpha$ depends on $\Gamma_0$. In supplement it is proven analytically that $N C_\alpha/\alpha$ for \emph{any} potential $U(\psi)$ in the PT limit $\alpha\ll1$. In our case, using $\psi_\mathrm{dec}=\arccos[-(2+\Gamma_0)/\Gamma_0]$, see Fig.~\ref{Fig:Pot}, and $\psi(E_\mathrm{max})=\pi$ we get
\begin{equation}
  C_\alpha^\mathrm{PT} = \int_{\psi_\mathrm{dec}}^{\pi} \fracp{U(\psi_\mathrm{st})}{\psi_\mathrm{st}}\frac{d\psi_\mathrm{st}}{W(\psi_\mathrm{st})}
  , \label{Eq:C_alpha.PT}
\end{equation}
where
\begin{equation}
  W(\psi_\mathrm{st}) = \int_{-\psi_\mathrm{st}}^{+\psi_\mathrm{st}} \sqrt{2[U(\psi_\mathrm{st})-U(\psi)]}\,d\psi
  . \label{Eq:W(psi)}
\end{equation}
The values of $C_\alpha$ obtained from direct simulations as well as $C_\alpha^\mathrm{PT}$ calculated using Eq.~\eqref{Eq:C_alpha.PT} are summarized in Tab.~\ref{Tab:KeyVals} together with other key numbers.

\begin{table}[!htb]
  \begin{tabular}{ddddd}
    \Gamma_0 & \gamma_{c-} & \alpha_R(\gamma_{c-}) & C_\alpha(\Gamma_0) \rem{& C_\gamma(\Gamma_0)} & C_\alpha^\mathrm{PT}(\Gamma_0)
    \\\hline
    -1.5 & 0.153 & 0.100 & 0.521 \rem{& 0.804} & 0.522\\
    -2.0 & 0.369 & 0.229 & 0.412 \rem{& 0.666} & 0.412\\
    -3.0 & 0.840 & 0.492 & 0.313 \rem{& 0.550} & 0.314\\
    -4.0 & 1.327 & 1.327 & 0.263 \rem{& 0.496} & 0.264\\
    \hline
  \end{tabular}
  \caption{
    The values of key quantities
  }
  \label{Tab:KeyVals}
\end{table}

\Sec{Conclusions}

In conclusion, we have stufied the retrapping of the phase in a point-like $\varphi$ JJ upon transition to zero-voltage state. For given damping $\alpha$, we have calculated the retrapping current $\gamma_R$ and the destination well, where the phase is trapped. For large $\alpha$ it is always a deeper well ($+\varphi$ for $\gamma>0$). However, as $\alpha$ decreases, the dependence of the destination well of $\alpha$ is an oscillating function, with oscillations (switching of the destination well) happening faster and faster even on the logarithmic scale, see Fig.~\ref{Fig:well-alpha}. Thus, at $\alpha\to0$ a tiny variation of $\alpha$ or $\gamma$ (noise) leads to a different destination well, \ie, to a butterfly effect. Detailed treatment of the noise will be presented elsewere.

%the possible consequences and applications of this effect
The butterfly effect at small damping does not allow to manipulate the $\varphi$ JJ by means of the bias current as described earlier\cite{Goldobin:CPR:2ndHarm,Sickinger:2012:varphiExp}. Simultaneously, in this regime one can use a $\varphi$ JJ as a random number generator (coin/dice) giving the output of $-\varphi$ or $+\varphi$ randomly. The extreme sensitivity may also be exploited in amplifiers or detectors as well as for the investigation of the fine details of the JJ dynamics itself. In the quantum regime the dynamics described here may lead to extremely strong mixing/entanglement of the states \ket{-\varphi} and \ket{+\varphi}.

\ifPRL\else
The butterfly effect is a \emph{necessary} condition for chaos. However chaos can occur in continuous flows of dimension $D\geq3$, \eg, in a driven system\cite{Yugay:2006:LJJ:Butterfly}. Our system is two-dimensional ($D=2$, dynamical variable being $\psi$ and $\dot{\psi}$), so that the chaos is impossible. To describe the chaotisity (and the butterfly effect) quantitatively one usually uses a Lyapunov exponent, which, in essence, the stability matrix averaged over the trajectory\cite{Scott:2005:ENLS,Sprott:2003:Chaos}. The Lyapunov exponent shows how fast two trajectories with a tiny difference in initial conditions diverge in the phase space. In our case, the Lyapunov exponents for any retrapping trajectory vanishes, which is obvious from the focuses on the phase plane, see Fig.~\ref{Fig:Traj}. Thus, a $\varphi$ Josephson junction is a an example of a non-chaotic system exhibiting a butterfly effect.
\fi

\acknowledgments
We acknowledge financial support by the DFG (via projects SFB/TRR-21 and GO~1106/3-1)
\ifPRL.\else
and by the German Israeli Foundation (Grant No. G-967-126.14/2007)
\fi

\bibliography{JJ,SF,NLP,QuComp}

\end{document}